\documentclass[%
 aip,
 amsmath,amssymb,floatfix,
 reprint,%
]{revtex4-1}
\usepackage[breaklinks=true,colorlinks=true,linkcolor=blue,urlcolor=blue,citecolor=blue]{hyperref}
\usepackage{graphicx}% Include figure files
\usepackage{dcolumn}% Align table columns on decimal point
\usepackage{bm}% bold math
\usepackage{changes}
\usepackage[utf8]{inputenc}
\usepackage[T1]{fontenc}
\usepackage{mathptmx}
\usepackage{etoolbox}
\usepackage{booktabs}
\usepackage{physics}
\usepackage{tabularx} 
\usepackage{xcolor}
\usepackage[]{units}
\usepackage[]{upgreek}

\DeclareUnicodeCharacter{0308}{\"}
\DeclareUnicodeCharacter{030C}{\v{ }}
\DeclareUnicodeCharacter{0301}{\'{ }}

\newcommand{\vect}[1]{\mathbf{#1}}
\UseRawInputEncoding 
\makeatletter
\def\@email#1#2{%
 \endgroup
 \patchcmd{\titleblock@produce}
  {\frontmatter@RRAPformat}
  {\frontmatter@RRAPformat{\produce@RRAP{*#1\href{mailto:#2}{#2}}}\frontmatter@RRAPformat}
  {}{}
}%
\makeatother
\begin{document}

\preprint{AIP/123-QED}
%\title{Surface acoustic wave effects in dc driven spintronic auto-oscillateur based magnetic vortex}
\title{Injection locking in  DC-driven spintronic vortex oscillators via surface acoustic wave modulation}
%\title{Non-linear behavior of current-driven magnetic vortices in the presence of SAW}

\author{R. Moukhader} \affiliation {Department of Mathematical and Computer Sciences, Physical Sciences and Earth Sciences, University of Messina, 98166, Messina, Italy}
\author{D. R. Rodrigues} \affiliation {Department of Electrical and Information Engineering, Politecnico di Bari, 70126 Bari, Italy}
\author{A. Riveros} \affiliation {Escuela de Ingenier\'ia, Universidad Central de Chile, 8330601 Santiago, Chile}
\author{A. Koujok} \affiliation{Fachbereich Physik and Landesforschungszentrum OPTIMAS, Rheinland-Pf\"alzische Technische Universit\"at Kaiserslautern-Landau, 67663 Kaiserslautern, Germany}
\author{G. Finocchio} \affiliation {Department of Mathematical and Computer Sciences, Physical Sciences and Earth Sciences, University of Messina, 98166, Messina, Italy}

\author{P. Pirro} \affiliation{Fachbereich Physik and Landesforschungszentrum OPTIMAS, Rheinland-Pf\"alzische Technische Universit\"at Kaiserslautern-Landau, 67663 Kaiserslautern, Germany}

\author{A. Hamadeh*} \affiliation{Fachbereich Physik and Landesforschungszentrum OPTIMAS, Rheinland-Pf\"alzische Technische Universit\"at Kaiserslautern-Landau, 67663 Kaiserslautern, Germany}
\email{hamadeh@rptu.de}

\date{\today}
             
\begin{abstract}

{Efficient control of spin-transfer torque oscillators (STOs) is crucial for their applications in non-volatile data storage, spin wave generation, and neuromorphic computing. This study investigates injection locking of a DC-driven vortex STO using surface acoustic waves (SAWs) to enhance the STO's signal, and allow for its synchronization with external inputs. We employ a simplified model based on Thiele's formalism, and highlight the role of vortex deformations in achieving injection locking. Micromagnetic simulations are conducted to validate our theoretical predictions, revealing how the locking bandwidth depends on SAW amplitude, as well as on the amplitude and direction of an applied external field. Our findings are pivotal for advancing experimental research and developing efficient low-power synchronization methods for large-scale STO networks.}

\end{abstract}

\maketitle

\section{Introduction}

Spintronic technology offers promising solutions to the critical challenges posed by rapid advances in artificial intelligence, particularly in enhancing scalability and reducing power consumption. \cite{hirohata2020review,dieny2020opportunities,finocchio2023roadmap} While spintronics has already seen industrial implementation, a key focus of ongoing research is the development of more efficient methods for manipulating and electrically reading magnetization dynamics.

The spin-transfer torque oscillator (STO) is a fundamental component in spintronic circuits. It is an intrinsically non-linear device with a three-layered structure in which a non-magnetic spacer separates two ferromagnetic layers. STOs offer notable frequency tunability through the use of electrical currents, magnetic fields, ambient noise, or interactions with other STOs.\cite{bonin2009analytical,dussaux2011phase,PhysRevB.85.140408,nakada2012noise,locatelli2015efficient,lebrun2017mutual,hamadeh2023}

STOs employing magnetic vortices have been demonstrated to generate signals in the sub-GHz range with high output powers \cite{belanovsky2012phase,hamadeh2014origin,locatelli2015efficient,hamadeh2024core} and narrow frequency linewidth. \cite{pribiag2007magnetic,ruotolo2009phase,locatelli2011dynamics}
This frequency range is typically associated with the vortex's gyrotropic mode, corresponding to a rotation of the core centered around its equilibrium position. The gyrotropic mode can be excited by dc currents above a characteristic threshold, determined by the geometry and material parameters, via spin-transfer torque (STT).\cite{yamada2007electrical,guslienko2008dynamic,nakano2011all}
Efficiently manipulating the gyrotropic dynamics of the magnetic vortex in STOs is of great interest due to the prominent applications it has to offer, such as non-volatile data storage \cite{park2003magnetic,sousa2005non} and spin wave generation. \cite{wintz2016magnetic,mayr2021spin,hamadeh2022hybrid} Generating non-linear effects in magnetic vortices, such as high frequencies and vortex core polarity reversal, typically requires high input currents. This results in increased Joule heating and ohmic losses that degrade energy efficiency. In contrast, voltage-driven magnetization excitation techniques can be more effective because they use electric fields rather than currents, thereby minimizing Joule heating and improving energy efficiency.\cite{bukharaev2018straintronics,mahmoud2020introduction}
%\textcolor{red}{The process of exciting magnetization dynamics directly via electric fields is rather challenging, as magnetic and electric fields exhibit weak coupling in most materials.}
%The process of exciting magnetization dynamics directly via electric fields is rather tedious.
One efficient way to excite magnetization dynamics using voltage gates is through hybrid magneto-electric methods.\cite{bukharaev2018straintronics}
This approach typically combines piezoelectric and magneto-elastic materials. When a voltage is applied, it induces mechanical deformation in the piezoelectric material, which interacts with the magneto-elastic material to generate magnetization dynamics through the Villari effect.\cite{tancrell1971wavefront,holland1974practical,maines1976surface,hashimoto2000surface,hess2002surface,weiler2011elastically,dreher2012surface,schmalz2020multi,9849071,ng2022excitation,koujok2023resonant,kunz2024coherent}
An interdigital transducer (IDT) on a piezoelectric substrate allows precise control of the mechanical deformations. The effective excitation of the gyrotropic mode of magnetic vortices by spin acoustic waves (SAW) has been demonstrated both theoretically\cite{koujok2023resonant} and experimentally.\cite{iurchuk2023excitation}

\begin{figure}[!t]
  \includegraphics[width=\columnwidth]{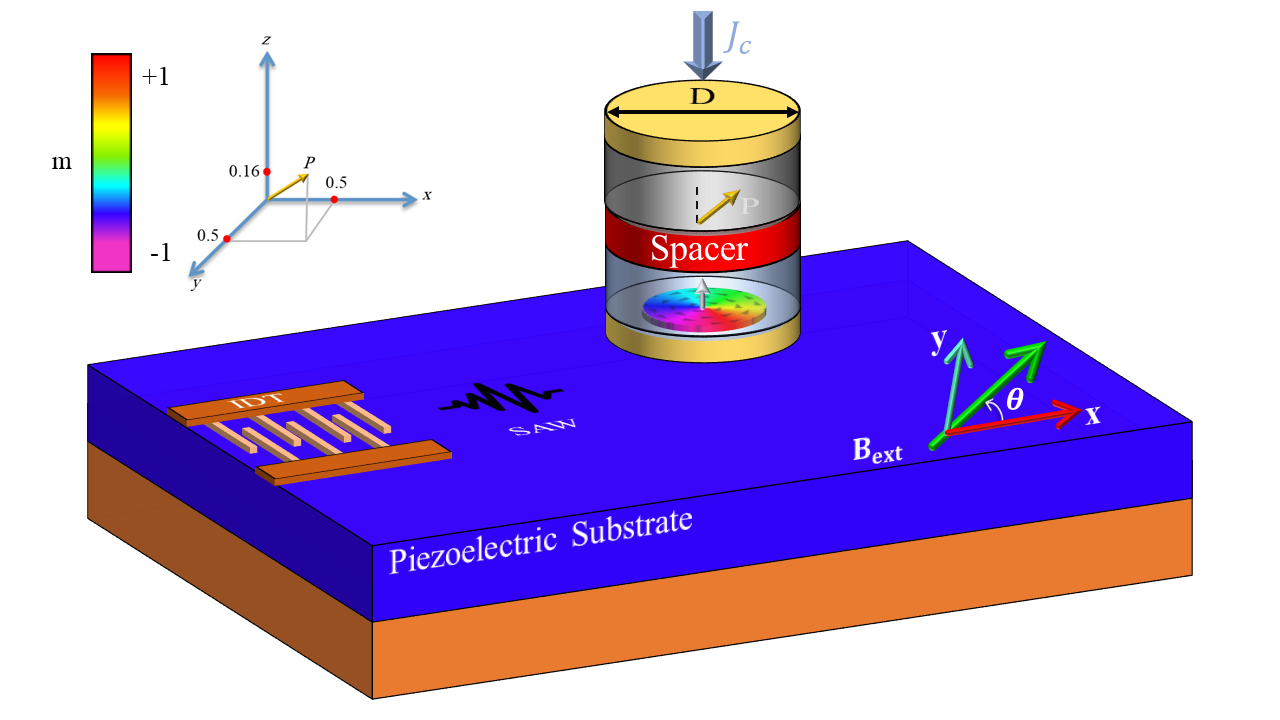}
  \caption{Schematic representation of the tri-layered pillar stack composed of a polarizer, a spacer and a free CoFeB layer in a magnetic vortex configuration. A DC current $J$ is injected via metallic electrodes through the fixed layer. The vortex STO is placed at a distance from the IDT which is used to excite propagating SAWs with surface localization. An external magnetic field $B_\mathrm{ext}$ is applied in the film's plane at an angle $\theta_B$ relative to the x-axis. (The figure does not accurately represent the relative sizes; it is not to scale)}
  \label{fig:1}
\end{figure}

To implement large networks of STOs, it's essential to synchronize their frequencies globally since variations between devices can cause frequency discrepancies. This work demonstrates theoretically that SAWs can control the frequency of the DC current-driven gyrotropic mode in vortex STOs.
FIG. \ref{fig:1} illustrates the multi-layered system considered. The three-layer column stack comprises a polarizer, a spacer, and a free Cobalt$-$Iron$-$Boron (CoFeB) layer in a vortex configuration. The disks are \unit[125]{nm} in diameter with a \unit[5]{nm} thick free layer. DC current is introduced into the vortex-based STO by means of metallic electrodes contacting the upper and lower ferromagnetic layers. 
These electrodes, which are only a few nanometers thick, are significantly smaller than the SAW wavelength which is in the micrometer range. This size difference ensures minimal disruption of SAW propagation, effectively preserving the strain effects on the magnetic vortex.
Incorporating these electrodes improves electrical contact, which is critical for efficient SAW generation and effective vortex-SAW coupling. Our predictions, based on the simplified Thiele approach and validated by micromagnetic simulations, suggest that this method can achieve injection locking. This work provides an avenue for the efficient and low-power synchronization of large STO networks.

\section{Analytical Model and Numerical Calculations}

To investigate the injection locking, we first use an effective approach based on the Thiele formalism. This method captures the essential dynamics of the magnetic vortex core under certain assumptions, and provides a simplified description compared to the more complex Landau-Lifshitz-Gilbert (LLG) equation. \cite{guslienko2001field, guslienko2008magnetic, gaididei2010magnetic,de2023quantitative,abreu2022ampere} By focusing on fewer degrees of freedom, the Thiele formalism helps clarify the fundamental principles of low-frequency dynamics. The resulting equations of motion describe the dynamics of the vortex core in terms of its position ($\vect{R}$) as follows,

%In the context of the Thiele formalism \cite{guslienko2001field,guslienko2008magnetic,gaididei2010magnetic}, the gyrotropic dynamics of a magnetic vortex can be described in terms of the VC's position. Employing the latter, the equations of motion describing the dynamics of the VC, in terms of its position ($\vect{R}$), are given by,

\begin{align}
	G[\dot{\vect{R}}\times\hat{\vect{z}}] = \vect{F}_{c} + \vect{F}_{nc} - \eta(R)\dot{\vect{R}}.
 \label{eq1}
\end{align}

Where $G = -2\pi  \Pi c$ is the gyroconstant which depends on the polarity $\Pi$ and the chirality $c$ of the vortex. In this study, we consider a counter clockwise chirality (c = -1), and a negative polarity ( $\Pi$ = -1), i.e. the vortex core points in the $-\hat{\vect{z}}$ direction.

The dissipation dyadic $\eta = \frac{ \alpha}{2} \int d^2 x \left( (\partial_x \mathbf{m})^2 + (\partial_y \mathbf{m})^2 \right) $ depends on the Gilbert damping $\alpha$, on the shape of the vortex and  the position $\mathbf{R}$ of the vortex core (VC) . $\vect{F}_{c} = - \frac{\gamma}{M_s }\partial_{\vect{R}}E$ is the net conservative force acting on the vortex, with $E$ being the free energy of the vortex. The non-conservative contribution $\vect{F}_{nc}$ arises due to the interaction of the vortex with the injected spin-polarized current. This force does not conserve energy and can drive the system out of equilibrium by acting as an anti-damping term. The non-conservative force due to STT is given by:
\begin{align}
	F_{STT\,i} = \sigma\int d^2 x \left(\vect{p}\cdot\left(\vect{m}\times\partial_{i}\vect{m}\right)\right),
 \label{eq2}
\end{align}
where $\sigma = g \mu_B J |P|/2eM_{s} h$, with $\mu_B$,$g$, $J$, $\vect{P}$, and $e$ being the Bohr magneton, Lande factor, the current density, the spin-polarization, and the electric charge, respectively. Moreover, $\vect{p} = \vect{P}/|\vect{P}|$ is the direction of the spin-polarization, corresponding to the orientation of the polarizer. The field-like torque term, which represents the conservative component of the spin-transfer torque, has been neglected because it does not qualitatively affect the vortex dynamics. However, our model includes the Oersted field resulting from the current injection, which contributes to the conservative force $\mathbf{F}_{c}$. This inclusion is essential for accurate modeling of vortex dynamics under applied currents.

The stability and dynamics of magnetic vortices, driven by exchange interactions, magnetostatic forces, STT, and Oersted fields, are well established in the literature.\cite{gaididei2010magnetic,de2023quantitative,abreu2022ampere} In this study, to predict injection locking using the Thiele formalism, it was crucial to incorporate the contribution of surface acoustic waves (SAW) into the free energy density. This contribution is given by
\begin{align}
	E_{mel} = B_{1}\epsilon_{XX}  \int d^2 x \; m_{x}^2.
 \label{eq3}
\end{align}

Here, we considered only the elastic tensor component $\epsilon_{XX}$ (longitudinal strain). The transverse strain components, $\epsilon_{YY}$ and $\epsilon_{ZZ}$ were neglected for simplicity. This approximation assumes that the longitudinal strain plays the dominant role in the magneto-elastic interaction under the conditions considered.
The parameter $B_\mathrm{1}$ represents the first-order magneto-elastic coupling constant. In this manuscript we define $\epsilon_{XX}(t) \equiv A_{SAW} \cos(\omega_{SAW} t)$, where $A_{SAW}$ and $\omega_{SAW}$ are the amplitude and frequency of the SAW, respectively. This strain component acts similarly to a unidirectional anisotropy along the x-direction. \cite{iurchuk2023excitation} This term can have two different effects on magnetic vortices. First, it can deform the core profile of a static vortex centered on the disk. Second, for vortices with displaced cores, it can drive gyrotropic motion. To displace the vortex core and break the spatial symmetry, thereby enhancing the magneto-elastic coupling, \cite{koujok2023resonant} we apply an in-plane external magnetic field at an angle of 45 degrees to the x-axis (see FIG. \ref{fig:1}).

While the Thiele formalism traditionally assumes a rigid magnetic texture, achieving injection locking requires accounting for changes in the vortex profile as it moves away from its equilibrium position. This non-linear behavior requires extending the dyadic dissipation to the second order in radial distance, going beyond the existing literature.\cite{guslienko2011spin}
The dissipation dyadic is expressed as 
\begin{align}
	\eta(\mathbf{R}) = \eta_0 + \eta_2 \mathbf{R}^2,
\label{eq4}
\end{align}
where $\eta_0 = \pi\alpha \ln(L/l)$ represents the zeroth-order dissipation with the vortex core at the disk's center, and $\eta_2 = \alpha \frac{\pi}{2 l^2}\left(1 - \left(\frac{l}{L}\right)^4 \right)$ is the second-order coefficient. Here, $l$ is the exchange length, and $L$ is the disk's radius.

\begin{figure}[!t]
  \centering
  \includegraphics[width=\columnwidth]{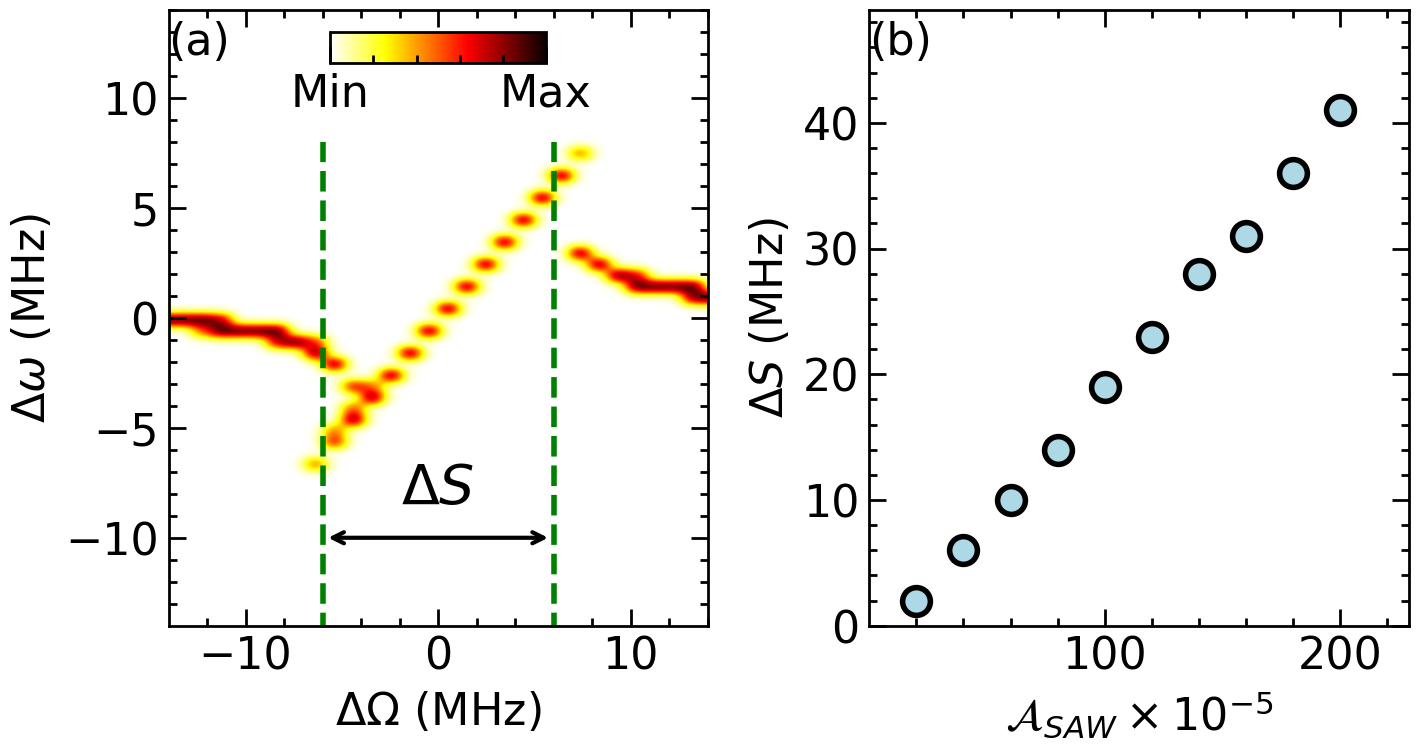}
 \caption{\textbf{Numerical results based on the Thiele model:} (a) Power spectral density (PSD) of the vortex frequencies excited by SAWs as a function of the SAW frequency. The locking bandwidth $\Delta S$ is indicated between the dashed green lines. (b) Variation of the locking bandwidth $\Delta S$ with respect to the SAW amplitude.}
  \label{fig:2}
\end{figure}

% Creating the table
\begin{table}[h]
    \caption{Parameters of CoFeB disk used in the micromagnetic simulations.}
    \label{table:example}
    \begin{tabular}{|c|c|c|}
        \hline
        Magnetic parameters & Symbol & Value\\ \hline
        Saturation Magnetization & $M_s$ & $1150 \times 10^{3}$ A/m \\ \hline
        Gilbert damping & $\alpha$ & 0.004 \\ \hline
        Exchange stifness & A & $15 \times 10^{-12}$ J/m \\ \hline
        Polarization & $\vect{P}=|P|\vect{p}$ & (0.5,0.5,0.16) \\ \hline
        First order magneto-elastic constant & $B_1$ & $-8 \times 10^{6}$ J/m$^{3}$  \\ \hline
        Uniaxial anisotropy constant & K& $2900$ J/m$^{3}$ \\ \hline
        Uniaxial anisotropy direction & $n_K$& (+1,0,0) \\ \hline
        Vortex chirality & $c$& -1 \\ \hline
        Vortex polarity & $\Pi$& -1 \\ \hline
    \end{tabular}
    \label{Table1}
\end{table}

To achieve injection locking using the Thiele formalism, we numerically integrated the model outlined in Eq. \ref{eq1}, incorporating the STT term from Eq. \ref{eq2}, the alternating magnetoelastic coupling from Eq. \ref{eq3}, and the second-order expansion of the dyadic dissipation from Eq. \ref{eq4}. This model also includes the established exchange and magnetostatic interactions, an external magnetic field, and the Oersted field. The analysis was performed with a DC current of \unit[3]{mA}, a SAW amplitude of $A_{SAW} = 60 \times 10^{-5}$, and a magnetic field of \unit[5]{mT} applied at a $\theta_B = 45^\circ$ angle to the $\hat{\vect{x}}$ direction (for more details about the parameters, see Table \ref{Table1}).
FIG.~\ref{fig:2}(a) shows a 2D plot of the power spectral density (PSD), illustrating the variation in the vortex STO's gyrotropic frequency as a function of the SAW's swept frequency. Here, $\Delta \omega = \omega - \omega_{g}$ represents the deviation of the oscillator frequency $\omega$ from its natural gyrotropic mode frequency $\omega_{g}$, and $\Delta \Omega = \omega_{SAW} - \omega_{g}$ represents the deviation of the SAW frequency $\omega_{SAW}$ from $\omega_{g}$. The plot highlights the locking bandwidth $\Delta S$, where frequency locking is achieved, and we also observe phase locking within this range.
FIG.~\ref{fig:2}(b) shows the locking bandwidth $\Delta S$ as a function of the SAW amplitude, revealing a linear increase in $\Delta S$ with the SAW amplitude within the studied range. These results confirm that by accounting for the non-linear dynamics of the vortex, injection locking in a current-driven vortex via SAW can be achieved, with the injection locking bandwidth tunable by the SAW amplitude. Importantly, in the absence of nonlinear effects, the locking bandwidth $\Delta S$ would be zero, highlighting the essential role of nonlinearity in enabling injection locking in this system. These are the main findings of this manuscript.

\section{Micromagnetic Simulations}
\begin{figure}
  \centering
  \includegraphics[width=\columnwidth]{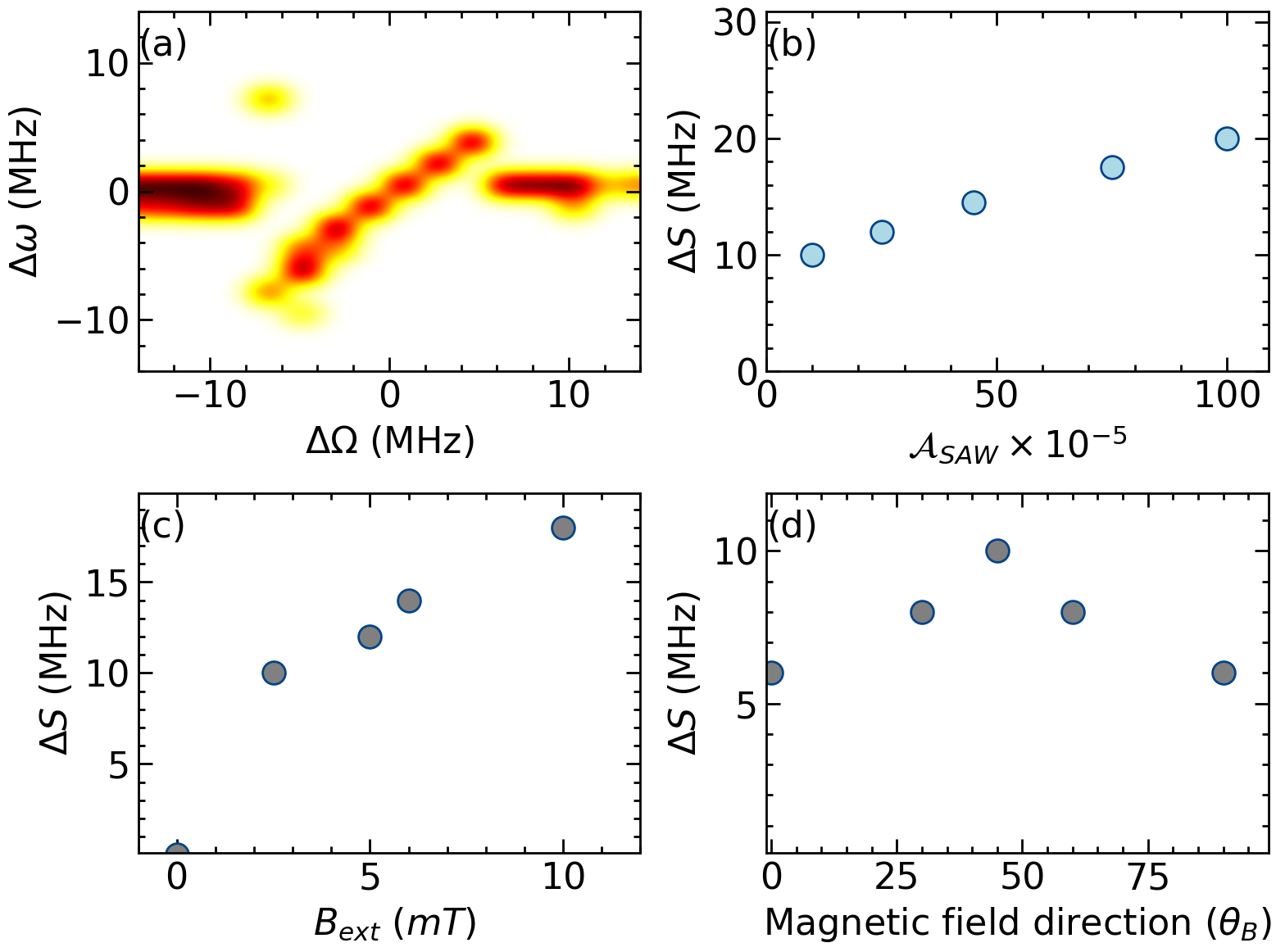}
 \caption{\textbf{Micromagnetic Simulations:} (a) PSD of the vortex frequencies excited by SAWs as a function of the SAW frequency. Variation of the locking bandwidth $\Delta S$ with respect to  (b) the SAW amplitude for fixed $B_{ext}$ and $\theta_B$, (c) the external magnetic field for fixed $A_{SAW}$ and $\theta_B$ and (d) the magnetic field's angle for fixed $B_{ext}$ and $A_{SAW}$.}

  \label{fig:3}
\end{figure}

To validate the predictions of the simplified Thiele model, we performed micromagnetic simulations using the full LLG equation. Unlike the Thiele approach which simplifies the system, micromagnetic simulations take into account all degrees of freedom, leading to variations in quantitative results especially in non-linear regimes. All parameters were consistent with those in the Thiele model, see table \ref{Table1}, except for the applied current $J$, which was set to \unit[7]{mA}. The simulations were performed with the Mumax3 software package \cite{vansteenkiste2014design} on the Aithericon platform. \cite{aithericon}

FIG.~\ref{fig:3} shows the results obtained from the micromagnetic simulations demonstrating the injection locking behavior. Panels (a) and (b) show that the locking behavior observed in these simulations is consistent with the predictions of the Thiele model. A fixed angle of $\theta_B = 45^\circ$ and $A_{SAW} = 10 \times 10^{-5}$ was used for these simulations. In addition, we investigated how the locking bandwidth depends on the amplitude of the magnetic field. 
FIG.~\ref{fig:3}(c) shows that the locking bandwidth increases linearly with the magnetic field amplitude, with simulations performed at a fixed angle of $\theta_B = 45^\circ$ and $A_{SAW} = 25 \times 10^{-5}$. This increase in bandwidth is attributed to the field-induced deformation of the magnetic vortex, which enhances the coupling with the SAW. As reported by Koujok \textit{et al.}\cite{koujok2023resonant}, the magnetic field breaks the symmetry of the vortex along the direction of the applied field, hence facilitating this coupling.
To further validate these findings, we examined the dependence of the locking bandwidth on the applied field angle $\theta_B$. FIG.~\ref{fig:3}(d) shows that the maximum bandwidth occurs at an angle of $\theta_B = 45^\circ$. In these simulations we considered a fixed field $B_{ext}$ = \unit[5]{mT} and $A_{SAW} = 10\times 10^{-5}$.

\begin{figure}[!t]
  \centering
  \includegraphics[width=\columnwidth]{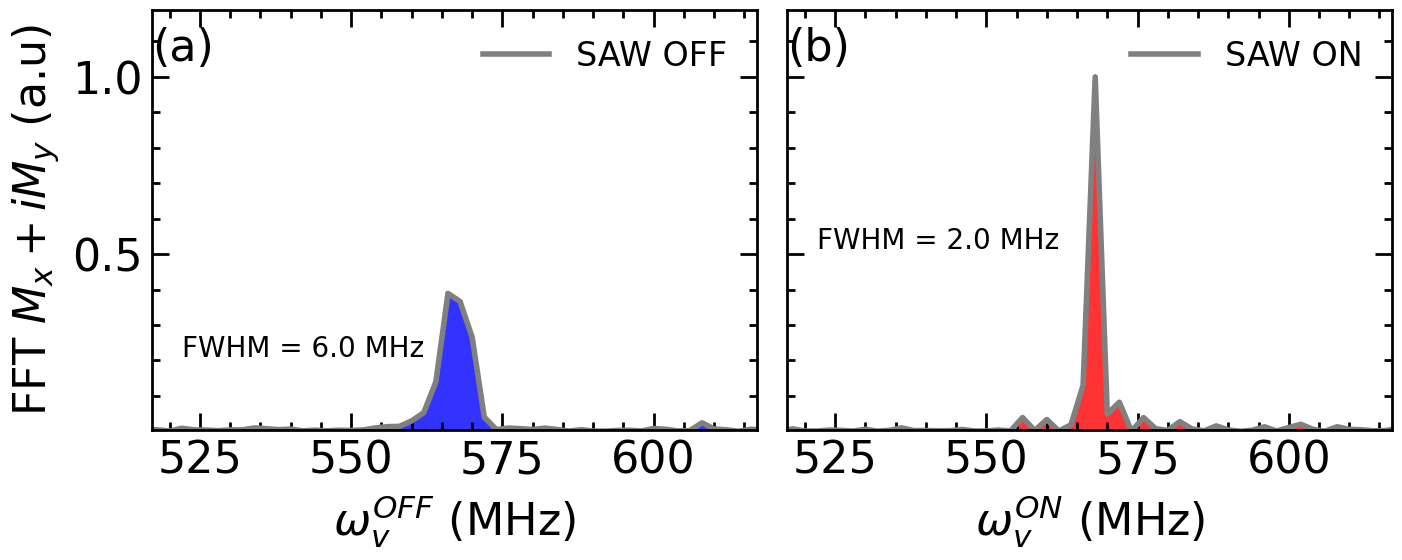}
  \caption{\textbf{Stochastic micromagnetic simulations:} FFT spectra of the magnetization dynamics at $B_{ext}$ = \unit[5]{mT}, $\theta_B = 45^\circ$ at room temperature (T = \unit[300]{K}) for two cases (a) in the absence of SAW and (b) in the presence of SAW with an amplitude $A_{SAW}$ = 1$\times 10^{-5}$.}
  \label{fig:4}
\end{figure}

To assess the impact of SAWs on the enhancement of signal properties under typical thermal conditions, we investigated their effect on the emitted signal of the vortex STO at a temperature of \unit[300]{K}. FIG.~\ref{fig:4} shows the Fast Fourier Transform (FFT) of the temporal magnetization components $M_x + iM_y$. In the absence of SAWs (FIG.~\ref{fig:4} (a), blue peak), the vortex STO's signal has a relatively low amplitude at the gyrotropic frequency. When SAWs with $A_{SAW} = 1 \times 10^{-5}$ are applied, a significant increase in the signal amplitude is observed (FIG.~\ref{fig:4} (b), red peak). This result is consistent with the locking behavior.\cite{lebrun2015understanding,hamadeh2014perfect}
Hence, injection locking via SAW enables efficient low-power linewidth narrowing, which is especially beneficial for applications requiring precise signal filtering or maintaining signal integrity over long distances.

\section{Summary and Conclusion}

In this study, we investigated the injection locking of DC current-driven vortex excitations by SAWs. Using the simplified Thiele formalism, we elucidated how vortex deformation induces non-linear behavior, and identified the resulting locking bandwidth and its dependence on SAW amplitude. These results were validated by micromagnetic simulations using the full LLG equation which confirmed the predicted behavior. Our simulations also revealed a linear relationship between the locking bandwidth and the applied magnetic field. In addition, analysis of the angle of the magnetic field further supported our understanding of the vortex deformation effects. At room temperature, we demonstrated that SAWs can significantly increase the oscillator signal amplitude while narrowing its bandwidth. These findings are critical for advancing experimental investigations and developing efficient low-power synchronization mechanisms for large-scale STO-based networks.

\begin{acknowledgments}
This work has been  supported  by the European union via the European Research Council within the Starting Grant No. 101042439 "CoSpiN" and by the Deutsche Forschungsgemeinschaft (DFG, German Research Foundation) - TRR 173 - 268565370" (project B01). DR and GF were supported by the project number 101070287 - SWAN-on-chip - HORIZON-CL4-2021-DIGITAL-EMERGING-01, the project PRIN 2020LWPKH7 "The Italian factory of micromagnetic modelling and spintronics" and the project PRIN20222N9A73 "SKYrmion-based magnetic tunnel junction to design a temperature SENSor-SkySens", funded by the Italian Ministry of University and Research (MUR) and by the PETASPIN Association (www.petaspin.com). DR also acknowledges funding from the project PE0000021, "Network 4 Energy Sustainable Transition - NEST", funded by the European Union - NextGenerationEU, under the National Recovery and Resilience Plan (NRRP), Mission 4 Component 2 Investment 1.3 - Call for Tender No. 1561 dated 11.10.2022 of the Italian MUR (CUP C93C22005230007) and the support of the project D.M. 10/08/2021 n. 1062 (PON Ricerca e Innovazione), funded by the Italian MUR. A.R. acknowledges the financial support from CIP2022036.

\end{acknowledgments}

\section*{Data Availability Statement}
The data underlying this study is available from the corresponding authors upon reasonable request.

\section*{Conflict of Interest}
The authors have no conflicts to disclose.

\section*{References}

\bibliography{bib}% Produces the bibliography via BibTeX.

\end{document}